\documentclass{eptcs}
\providecommand{\event}{SYNTCOMP 2018} % Name of the event you are submitting to
\usepackage{amsthm}
\usepackage{amssymb}
\usepackage{mathtools}
\usepackage{temporal}
\usepackage{enumerate} 
\usepackage{color}
\usepackage{tikz}
\usepackage{paralist}
\usepackage{fixme,manfnt,mparhack}
\usepackage[T1]{fontenc}
\usepackage{enumerate}
\usepackage{xspace}
\usepackage{amssymb}
\usepackage{paralist}
\usepackage{wrapfig}
\usepackage{bbm}
\usepackage{cite}
\usepackage{booktabs}
\usepackage{array}
\usepackage{arydshln}

\usepackage{macros}

\providecommand{\thisvolume}[1]{this volume of {\sl Electronic
  Proceedings in Theoretical Computer Science}}

\definecolor{mygreen}{rgb}{0,0.6,0} 
\definecolor{mygray}{rgb}{0.9,0.9,0.9} 
\definecolor{mymauve}{rgb}{0.58,0,0.82}

\usepackage{listings}
\usepackage{xspace}

\newcommand{\myparagraph}[1]{\smallskip \noindent {\em #1.}}

\newcommand{\available}[1]{is available at \url{#1}. Accessed April 2019.}

\newcommand{\demiurge}{\textsf{Demiurge}\xspace}

\newcommand{\syntcomp}{SYNTCOMP\xspace}

\newcommand{\abssynthe}{AbsSynthe\xspace}
\newcommand{\simpleBDD}{Simple BDD Solver\xspace}

\newcommand{\termitesat}{TermiteSAT\xspace}

\newcommand{\bowser}{BoWSer\xspace}
\newcommand{\ltlsynt}{\texttt{ltlsynt}\xspace}
\newcommand{\bosy}{BoSy\xspace}
\newcommand{\party}{\texttt{sdf-hoa}\xspace}
\newcommand{\partyold}{\textsc{Party} (aiger)\xspace}

\newcommand{\safetysynth}{SafetySynth\xspace}

\newcommand{\Abc}{\textsf{ABC}\xspace}

\newcommand{\subby}{\kern-0.2em\leftarrow\kern-0.2em}

\newcommand{\strix}{\textsc{Strix}\xspace}

\newcommand{\lazysynt}{\textsc{LazySynt}\xspace}

\title{The 5th Reactive Synthesis Competition (\syntcomp 2018): Benchmarks, Participants \& Results}
\author{Swen Jacobs
\institute{CISPA Helmholtz Center for Information Security\\Saarbr\"ucken, Germany}
\and
Roderick Bloem
\institute{Graz University of Technology \\ Graz, Austria}
\and 
Maximilien Colange
\institute{LRDE, EPITA\\ Kremlin-Bic\^etre, France}
\and 
Peter Faymonville
\institute{Saarland University\\Saarbr\"ucken, Germany}
\and 
Bernd Finkbeiner
\institute{Saarland University\\Saarbr\"ucken, Germany}
\and
Ayrat Khalimov
\institute{Hebrew University\\Jerusalem, Israel}
\and 
Felix Klein
\institute{Saarland University\\Saarbr\"ucken, Germany}
\and
Michael Luttenberger
\institute{Technical University of Munich\\Munich, Germany}
\and
Philipp J. Meyer
\institute{Technical University of Munich\\Munich, Germany}
\and
Thibaud Michaud
\institute{LRDE, EPITA\\ Kremlin-Bic\^etre, France}
\and
Mouhammad Sakr
\institute{CISPA Helmholtz Center for Information Security\\Saarbr\"ucken, Germany}
\and
Salomon Sickert
\institute{Technical University of Munich\\Munich, Germany}
\and
Leander Tentrup
\institute{Saarland University\\Saarbr\"ucken, Germany}
\and 
Adam Walker
\institute{Independent Researcher}
}
\def\titlerunning{\syntcomp 2018}
\def\authorrunning{S. Jacobs et al.}
\begin{document}
\maketitle

\begin{abstract}
We report on the fifth reactive synthesis competition (\syntcomp 2018). We introduce four new benchmark classes that have been added to the SYNTCOMP library, and briefly describe the evaluation scheme and the experimental setup of \syntcomp 2018. We give an overview of the participants of \syntcomp 2018 and highlight changes compared to previous years. 
Finally, we present and analyze the results of our experimental evaluation, including a ranking of tools with respect to quantity and quality of solutions.
\end{abstract}

\section{Introduction}
\label{sec:intro}
The synthesis of reactive systems from formal specifications, as first defined by Church~\cite{Church62}, is one of the major challenges of computer science. Until recently, research focused on theoretical results, with little impact on the practice of system design. Since 2014, the reactive synthesis competition (\syntcomp) strives to increase the practical impact of these theoretical advancements~\cite{SYNTCOMP14}. \syntcomp is designed to foster research in scalable and user-friendly implementations of synthesis techniques by establishing a standard benchmark format, maintaining a challenging public benchmark library, and providing an independent platform for the comparison of tools under consistent experimental conditions.
	
%Competitions have been a means of achieving these goals in many related
%fields, including automated reasoning~\cite{SutcliffeS06,BarrettMS05,JarvisaloBRS12}
%and verification~\cite{HWMCC,Beyer12}.
%In parallel to \syntcomp 2014, the
%\emph{syntax-guided synthesis competition} (SyGuS-COMP) was held for the first
%time.\cite{AlurBJMRSSSTU13} The focus of SyGuS-COMP is on the synthesis of
%functional instead of reactive programs, and the specification is given as a
%first-order logic constraint on the function to be synthesized, along with a syntactic
%constraint that restricts how solutions can be built.
%A significant difference of synthesis competitions to most of the competitions
%in automated reasoning or verification is that
%solutions to the synthesis problem can be ranked according to inherent quality
%criterions, such as reaction time or size of the solution. In addition to
%efficiency and correctness, a synthesis competition also needs to measure
%quality. 

\syntcomp is held annually, and competition results are presented at the International Conference on Computer Aided
Verification (CAV) and the Workshop on Synthesis (SYNT).~\cite{SYNTCOMP14,SYNTCOMP15,SYNTCOMP16,SYNTCOMP17} This year, like in its inaugural edition, \syntcomp was part of the FLoC Olympic Games. 
For the first two competitions, \syntcomp was limited to safety properties specified as monitor circuits in an extension of the AIGER format~\cite{SYNTCOMP-format}. \syntcomp 2016 introduced a new track that is based on properties in full linear temporal logic (LTL), given in the \emph{temporal logic synthesis format} (TLSF)~\cite{JacobsB16,JacobsK16}.
	
The organization team of \syntcomp 2018 consisted of R. Bloem and S. Jacobs.

\vspace{-5pt}
\paragraph{Outline.} In Section~\ref{sec:benchmarks}, we present four new benchmark classes that have been added to the \syntcomp library for \syntcomp 2018. In Section~\ref{sec:setup}, we briefly describe the setup, rules and execution of the competition, followed by an overview of the participants in Section~\ref{sec:participants}. Finally, we provide the experimental analysis in Section~\ref{sec:results} and concluding remarks in Section~\ref{sec:conclusions}.

\section{New Benchmarks}
\label{sec:benchmarks}

In this section, we describe benchmarks added to the \syntcomp library since the last iteration of the competition. For more details on the complete benchmark library, we refer to the previous competition reports~\cite{SYNTCOMP14,SYNTCOMP15,SYNTCOMP16,SYNTCOMP17}.

\subsection{Benchmark Set: Temporal Stream Logic (TSL)}
\label{sec:benchmarks-tsl}

This set of benchmarks stems from work by Finkbeiner et al.~\cite{FKPS17} on the synthesis of functional reactive programs (FRPs) that handle data streams. They are originally specified in temporal stream logic (TSL) and are not parameterized. The benchmark set covers:
\begin{itemize}
\item toy examples like simple buttons or incrementors, 
\item different versions of an escalator controller, 
\item a music app based on the Android music player library, 
\item a simple game where the player has to react to the movement of a slider, 
\item controllers for simulated autonomous cars in the \emph{The open racing car simulator} (TORCS) framework~\cite{torcs}, and
\item three benchmarks based on the FRP Zoo project~\cite{frpzoo} that specify the behavior of a program with buttons, a counter, and a display.
\end{itemize}

The set contains $24$ benchmarks that have been translated to TLSF by F. Klein and M. Santolucito.

\subsection{Benchmark Set:  Hardware Components}
\label{sec:benchmarks-components}

This benchmark set comprises several standard hardware components, such as a mux, an $n$-ary latch, a shifter, several variants of a collector that signals if all of its clients have sent a specific signal, as well as several new arbiter variants. 

The set contains $11$ parameterized benchmarks, encoded in TLSF by F. Klein.

\subsection{Benchmark: Tic Tac Toe}
\label{sec:benchmarks-ttt}

An infinite variant of Tic Tac Toe with repeated games, where each can end in a win, loss, or draw. The system player has to ensure that they never lose a game. This benchmark is not parameterized and has been encoded in TLSF by F. Klein. 

\subsection{Benchmark Set: Abstraction-based Control}
\label{sec:benchmarks-abstraction}

This set of benchmarks is based on the research on abstraction-based control synthesis by Nilsson, Liu and Ozay~\cite{NilssonOL17}. They are based on an abstraction of physical systems governed by ordinary differential equations to finite-state transition systems. The benchmark set contains a toy example and four variants of a linear inverted pendulum model. 

In preliminary experiments, the organizers detected that some of these examples are so big that already the translation from TLSF to different other formats takes more than 20 min. Therefore, these benchmarks have not been used in \syntcomp 2018, and will instead be posted as challenge benchmarks for the community to encourage further development of algorithms and solvers.

These benchmarks have been encoded into TLSF by Z. Liu.

\section{Setup, Rules and Execution}
\label{sec:setup}

We give a brief overview of the setup, rules and execution of \syntcomp 2018. Up to the selection of benchmarks for the LTL (TLSF) track, all of this is identical to the setting of \syntcomp 2017 (cf. the report of \syntcomp 2017~\cite{SYNTCOMP17} for more details).

\paragraph{General Rules.}
The competition has two main tracks, one based on safety specifications in AIGER format, and the other on full LTL specifications in TLSF. The tracks are divided into subtracks for \emph{realizability checking }and \emph{synthesis}, and into two execution modes: \emph{sequential} (using a single core of the CPU) and \emph{parallel} (using up to $4$ cores). 

Every tool can run in up to three configurations per subtrack and execution mode. Before the competition, all tools are tested on a small benchmark set, and authors can submit bugfixes if problems are found.

\paragraph{Ranking Schemes.}
In all tracks, there is a ranking based on the number of correctly solved problems within a $3600$s timeout. In the synthesis tracks, correctness of the solution additionally has to be confirmed by a model checker. 

Furthermore, in synthesis tracks there is a ranking based on the \emph{quality} of the solution, measured by the number of gates in the produced AIGER circuit. To this end, the size of the solution is compared to the size $ref$ of a reference solution. A correct answer of size $ref$ is rewarded $2$ points, and smaller or larger solutions are awarded more or less points, respectively.

\paragraph{Selection of Benchmarks.}
Benchmarks are selected according to the same scheme as in previous years, based on a categorization into different classes. For the safety (AIGER track), $234$ benchmarks were selected --- the same number of benchmarks per class, but a different set of benchmarks unless all benchmarks from a class are selected. For the LTL (TLSF) track, $286$ benchmark instances were selected. This includes $34$ parameterized benchmarks, each in up to $6$ instances.\footnote{Out of the $11$ new parameterized benchmarks, $10$ were selected. Only the \texttt{collector\_enc} benchmark was not used since it contains a ``Preset'' block that is not supported by all competing tools.}

\paragraph{Execution.}
Like in the last three years, \syntcomp 2018 was run at Saarland University, on a set of identical machines with a single quad-core Intel Xeon processor (E3-1271 v3, 3.6GHz) and 32 GB RAM (PC1600, ECC). Benchmarking was again organized on the EDACC 
platform~\cite{BalintDGGKR11}. 

The model checker for checking correctness of solutions for the safety (AIGER) track is IIMC\footnote{IIMC used to be available at \url{ftp://vlsi.colorado.edu/pub/iimc/}, but is currently unavailable as of April 2019.} in version 2.0, preceded by an invariant check\footnote{Available from the \syntcomp repository at \url{https://bitbucket.org/swenjacobs/syntcomp/src}, in subdirectory \texttt{tools/WinRegCheck}. Accessed April 2019.} for solvers that supply an inductive invariant in addition to their solution. 
For the LTL (TLSF) track, the model checker used was V3\footnote{V3 is available at \url{https://github.com/chengyinwu/V3}. Accessed April 2019.}~\cite{WuWLH14}.

\section{Participants}
\label{sec:participants}
Ten tools participated competitively in \syntcomp 2018: five in the safety track, and five in the LTL track. In addition, two tools were entered hors concours in the safety track. We briefly describe the participants and give pointers to additional information.
For additional details on the implemented techniques and optimizations, we refer to the previous \syntcomp reports~\cite{SYNTCOMP14,SYNTCOMP15,SYNTCOMP16,SYNTCOMP17}. 

\subsection{Safety Track}

Four of the five competitive participants in this track ran in the same version as in \syntcomp 2017. The only one that received an update is \simpleBDD. One additional tool was entered hors concours by M. Sakr and competition organizer S. Jacobs. 

\subsubsection{Updated Tool: \simpleBDD}

An update of \simpleBDD was submitted by A. Walker, and competed in the realizability track.
\simpleBDD implements the classical BDD-based fixpoint algorithm for safety games. In sequential mode, it runs in three configurations, two of which are based on an abstraction-refinement approach inspired by de Alfaro and Roy~\cite{dealfaro} (configurations abs1 and abs2), and one without any abstraction. All three implement many important optimizations. These configurations are the same as last year.

Additionally, three new configurations have been entered for the parallel mode, which run different portfolios of the algorithms in the sequential mode.

\paragraph{Implementation, Availability}
The source code of \simpleBDD is available online at
\url{https://github.com/adamwalker/syntcomp}. 

\subsubsection{Re-entered: Swiss \abssynthe v2.1} 
\abssynthe was submitted by R. Brenguier, G. A. P\'erez, J.-F. Raskin, and O. Sankur, and competed in the realizability and the synthesis tracks. It also implements the classical BDD-based algorithm for safety games, and additionally supports decomposition of the problem into independent sub-games, as well as an abstraction approach~\cite{BrenguierPRS14,BrenguierPRS15,SYNTCOMP17}. 
It competes in its best-performing configurations from last year:
\begin{itemize}
\item sequential configuration (SC2) uses abstraction, but no compositionality,
\item sequential configuration (SC3) uses a compositional algorithm, combined with an abstraction method, and 
\item parallel configuration (PC1) is a portfolio of configurations with different combinations of abstraction and compositionality.
\end{itemize}

\paragraph{Implementation, Availability}
The source code of \abssynthe is is available at
\url{https://github.com/gaperez64/AbsSynthe/tree/native-dev-par}.

\subsubsection{Re-entered: \demiurge 1.2.0} 
\demiurge was submitted by R. K\"onighofer 
%from Graz University of  Technology 
and M. Seidl,
%from Johannes-Kepler-University Linz, 
and competed in the realizability and the synthesis tracks.
\demiurge implements SAT-based methods for solving safety games~\cite{BloemKS14,SeidlK14}. In the competition, three different methods are used --- one of them as the only method in sequential configuration (D1real), and a combination of all three methods in parallel configuration (P3real). This year, \demiurge competed in the same version as in previous years. 

\paragraph{Implementation, Availability}
{ \sloppy The source code of \demiurge is
available at {\url{%
https://www.iaik.tugraz.at/content/research/opensource/demiurge/}}
under the GNU LGPL version 3. }

\subsubsection{Re-entered: \safetysynth}
\safetysynth was submitted by L. Tentrup, % from Saarland University, Saarbr\"ucken
and competed in both the realizability and the synthesis track. \safetysynth implements the classical BDD-based algorithm for safety games, using the optimizations that were most beneficial for BDD-based tools in \syntcomp 2014 and 2015~\cite{SYNTCOMP14,SYNTCOMP15}. It competed in the single most successful configuration from last year.

\paragraph{Implementation, Availability}
The source code of \safetysynth is available online at: \url{https://www.react.uni-saarland.de/tools/safetysynth/}.

\subsubsection{Re-entered: \termitesat}
\termitesat was submitted by A. Legg, N. Narodytska and L. Ryzhyk, and competed in the realizability track (parallel mode). \termitesat implements a SAT-based method for solving safety games based on Craig interpolation. It competes in its most successful configuration from last year, which implements a hybrid method that runs this algorithm alongside one of the algorithms of \simpleBDD~\cite{LeggNR16}, with communication of intermediate results between the different algorithms.

\paragraph{Implementation, Availability}
The source code of \termitesat is available online at: \url{https://github.com/alexlegg/TermiteSAT}.
%

%\subsubsection{Hors Concours: Simple SDD Solver}\todo{remove Simple SDD?}
%Simple SDD Solver has been submitted by A. Walker and is a port of \simpleBDD that uses Sentential Decision Diagrams (SDDs)~\cite{Darwiche11} instead of BDDS.
%
%\paragraph{Implementation, Availability}
%The source code of Simple SDD Solver is available online at: \url{https://github.com/adamwalker/syntcomp-sdd}.

\subsubsection{Hors Concours: \lazysynt\ --- Symbolic Lazy Synthesis}
\lazysynt has been submitted by M. Sakr and S. Jacobs. It participated hors concours in the synthesis track. In contrast to the classical BDD-based algorithm and the SAT-based methods implemented in \demiurge, \lazysynt implements a combined forward-backward search that is embedded into a refinement loop, generating \emph{candidate} solutions that are checked and refined with a combination of backward model checking and forward generation of additional constraints.~\cite{JacobsS18} 

\paragraph{Implementation, Availability}
\lazysynt is implemented in Python and uses the CUDD library for BDD operations and \Abc for compression of AIGER solutions. 

\subsection{LTL-Track}
\label{sec:participants-TLSF}

This track had five participants in 2018, one of which is a new entrant that has not participated in previous competitions. Four tools competed in both the realizability and the synthesis track, one only in the realizability track. We briefly describe the synthesis approach and implemented optimizations of the new tool, followed by an overview of the changes in re-entered tools. For additional details on the latter, we refer to the reports of \syntcomp 2016 and 2017~\cite{SYNTCOMP16,SYNTCOMP17}.

\subsubsection{New Entrant: \strix}
\strix was submitted by P. J. Meyer, S. Sickert and M. Luttenberger. It competed in both the realizability and the synthesis track.

\strix implements a translation of the LTL synthesis problem to parity games. At its core, it uses explicit-state game solving based on strategy iteration~\cite{BjorklundSV04,Schewe08,Luttenberger08}. To make the approach efficient, it uses two main optimizations:
\begin{itemize}
\item The specification is decomposed into smaller parts that can be more efficiently translated to an automata representation. This usually results in a large number of simple safety and co-safety conditions, and only few conditions that require automata with more complicated acceptance conditions. Moreover, this step also detects and exploits symmetry in the specification.
\item Game solving begins before the full specification is translated, and missing parts of the specification are only translated into the automata representation if they are relevant for solving the game.
\end{itemize}

\strix competed in the following configurations:
\begin{itemize}
\item one configuration each for the sequential and the parallel realizability tracks (with the only difference being the number of running threads),
\item three configurations for the sequential synthesis track, which only differ in the construction of the AIGER solution from the Mealy machine that is constructed as an intermediate result: configuration \texttt{basic} directly encodes the Mealy machine into AIGER, configuration \texttt{min} minimizes the number of states in the Mealy machine before translating it to AIGER, and configuration \texttt{labels} uses a labeling of states in the Mealy machine based on the product automaton, and assigns a separate set of latches in the AIGER circuit to each component of the product. 
\item in parallel synthesis mode, \strix invokes the three sequential synthesis configurations in parallel, as well as an additional method that combines the approaches of \texttt{min} and \texttt{labels}.
\end{itemize}

\paragraph{Implementation, Availability} \strix is implemented in Java and C++. It uses \textsc{MeMin}\footnote{\textsc{MeMin} \available{https://embedded.cs.uni-saarland.de/MeMin.php}}~\cite{AbelR15} for the minimization of Mealy machines constructed from winning strategies, and \Abc\footnote{\Abc \available{https://github.com/berkeley-abc/abc}}~\cite{BraytonM10} for the compression of AIGER circuits obtained from these Mealy machines. 

The source code of \strix is available at \url{https://strix.model.in.tum.de/}.

\subsubsection{Updated Tool: \bowser}
\bowser was submitted by B. Finkbeiner and F. Klein. It implements different 
extensions of the bounded synthesis approach~\cite{Finkbeiner13} 
that solves the LTL synthesis problem by first translating the specification 
into a universal co-B\"uchi automaton, and then encoding acceptance of a 
transition system with bounded number of states into  a constraint system, in 
this case a propositional satisfiability (SAT) problem. A solution to this 
SAT problem then represents a transition system that satisfies the original 
specification. To check for unrealizability of the specification, the 
existence of a winning strategy for the environment is also encoded into SAT. 

In the basic configuration, a solution from the SAT solver is directly 
encoded into an AIGER circuit, and then handed to Yosys for simplification. 
As extensions, \bowser implements 
\emph{bounded cycle synthesis}~\cite{FinkbeinerK16}, which restricts the 
structure of the solution with respect to 
the number of cycles in the transition system, as well as a third encoding 
that puts bounds on the number of gates and latches in the resulting AIGER 
circuit.

Compared to last year's version, a number of small improvements to speed up computations have been implemented, and an experimental preprocessor for LTL formulas has been added, and is now used in configuration (synth), see below.

\bowser competed in sequential and parallel variants of the following configurations:
\begin{itemize}
\item configuration (basic) implements bounded synthesis in the basic version mentioned above,
\item configuration (synth) implements the same approach as (basic), except that it uses an additional preprocessor for LTL formulas to simplify the specification. bounded cycle synthesis on top of bounded synthesis, 
\item configuration (opt) also implements bounded cycle synthesis on top of bounded synthesis, i.e., in a first step it searches for a solution with a bounded number of states, and if that exists, it additionally bounds the number of cycles. 
\end{itemize}

In sequential mode, these configurations spawn multiple threads that are executed on a single CPU
core. The parallel configurations are mostly the same as the sequential ones, but use a slightly different strategy for exploring the search space of solutions.

\paragraph{Implementation, Availability}
\bowser is implemented in Haskell, and uses Spot\footnote{Spot \available{https://spot.lrde.epita.fr}} to convert specifications into automata, and MapleSAT\footnote{MapleSAT \available{https://sites.google.com/a/gsd.uwaterloo.ca/maplesat/}}~\cite{LiangGPC16} to solve SAT queries. For circuit generation, it uses the Yosys framework\footnote{Yosys \available{http://www.clifford.at/yosys/}}~\cite{Glaser2014}.
The website of \bowser is \url{https://www.react.uni-saarland.de/tools/bowser/}.

\subsubsection{Updated Tool: \ltlsynt}
\ltlsynt was submitted by M. Colange and T. Michaud and competed in three different configurations in both the sequential realizability and sequential synthesis tracks.
 
To solve the synthesis problem, \ltlsynt uses a translation to parity games. To increase efficiency, it uses an intermediate translation to $\omega$-automata with a \emph{transition-based} generalized B\"uchi acceptance condition and simplified based on heuristics.~\cite{Duret16}
To convert the automaton into a game, transitions are split into two separate actions, one action of the environment and one action of the controller, and the resulting automaton is finally determinized and can then be interpreted as a parity game (with transition-based winning condition).
To solve this parity game, \ltlsynt uses the well-known recursive algorithm by Zielonka~\cite{Zielonka98}, adapted to games with transition-based winning condition. 
A winning strategy for the parity game defines a satisfying implementation of the controller in the synthesis problem. \ltlsynt encodes the strategy into an AIGER circuit using an intermediate representation in Binary Decision Diagrams (BDDs), allowing some simplifications. 
%In contrast to its competitors, \ltlsynt does not use external tools such as \Abc or Yosys for the encoding of solutions into AIGER.

Compared to last year's version, \ltlsynt uses additional optimizations in the determinization and in Zielonka's algorithm, and it now competes in three different configurations:
\begin{itemize}
\item configuration (sd) implements splitting before determinization (as described above),
\item configuration (ds) implements splitting after determinization, and
\item configuration (incr) implements an incremental form of determinization.
\end{itemize}

\paragraph{Implementation, Availability}
%\todo{taken from last year, is this still correct? new dependencies?}
\ltlsynt is implemented in C++ and integrated into a branch of the Spot automata library~\cite{Duret16}, which is used for translation of the specification into automata, and for manipulation of automata. Spot also integrates the BDD library BuDDy and the SAT solver PicoSAT. 

The source code of \ltlsynt is available in branch \texttt{tm/ltlsynt-pg} of the GIT repository of Spot at \url{https://gitlab.lrde.epita.fr/spot/spot.git}.

\subsubsection{Updated Tool: \bosy} 
\bosy was submitted by P. Faymonville, B. Finkbeiner and L. Tentrup, and competed in both the realizability and the synthesis track. 
To detect realizability, \bosy translates the LTL specification into a co-B\"uchi automaton, and then to a safety automaton by bounding the number of visits to rejecting states.
The resulting safety game is solved by \safetysynth.
For synthesis, \bosy additionally implements \emph{bounded synthesis}~\cite{Finkbeiner13} with an encoding into quantified Boolean formulas (QBF)~\cite{FaymonvilleFRT17,FaymonvilleFT17}. 
%One advantage of this encoding is that it allows to keep the input valuations symbolic, and to solve the constraints without explicitly enumerating all possibilities (as existing SMT solvers do). 
%To obtain small solutions, \bosy does not require solutions to be input-preserving (cp. Finkbeiner and Schewe~\cite[Sect. 8.4]{Finkbeiner13}).
To detect unrealizability, the existence of a winning strategy of the environment is encoded in a similar way and checked in parallel. 
The resulting QBF formulas are simplified using the QBF preprocessor bloqqer11~\cite{BiereLS11}. To solve the QBF constraints, \bosy uses a combination of CAQE~\cite{RabeT15,Tentrup17} and the certifying QBF solver QuAbS~\cite{Tentrup16}, and the certificate returned by QuAbS represents a solution to the synthesis problem. This solution is then converted into an AIGER circuit, and further simplified using the \Abc framework. 
The resulting strategy (if any) is then compared to the solution found by \safetysynth, and the smaller one is returned. 

Two configurations of \bosy competed in \syntcomp 2018: configuration (basic) and configuration (opt), where the latter further improves the size of the strategy by encoding the existence of an AIGER circuit representing the strategy directly into a QBF query. Both configurations support a parallel mode, if more than one core is available. 

\paragraph{Implementation, Availability.}
\bosy is written in Swift. It uses LTL3BA\footnote{LTL3BA \available{https://sourceforge.net/projects/ltl3ba/}}~\cite{ltl3ba} or Spot to convert LTL specifications into B\"uchi automata. It uses bloqqer11\footnote{Bloqqer \available{http://fmv.jku.at/bloqqer}}, CAQE\footnote{RAReQS \available{https://www.react.uni-saarland.de/tools/caqe/}} and QuAbs\footnote{QuAbs \available{https://www.react.uni-saarland.de/tools/quabs/}} to solve QBF constraints, and \Abc to simplify solutions.

The code is available online at: \url{https://www.react.uni-saarland.de/tools/bosy/}.

\subsubsection{Updated Tool: \party (previously \partyold)}

\party was submitted by A. Khalimov, and competed in the sequential realizability track in a single configuration. The tool is a C++ rewrite with minor optimizations of the aiger configuration of last year's entrant \partyold~\cite{KhalimovJB13}.

\party follows a variant of the bounded synthesis approach~\cite{Finkbeiner13}
 and translates a given specification to a universal safety automaton that 
approximates liveness properties by bounded liveness. The transition relation 
of the safety automaton is translated into a symbolic representation with 
BDDs, and then treated as a safety game. The game is solved using the 
standard symbolic fix-point algorithm.

Compared to the last year's \partyold, \party removes the unnecessary 
step in the translation "automaton->aiger->bdd", which was a technical 
detail, and goes directly "automaton->bdd".

\paragraph{Implementation, Availability.}
\party is written in C++. As input, it accepts a universal co-B\"uchi automaton in 
HOA format, the output is a Yes/No answer and no circuit.
For the competition, a wrapper script converted a given LTL formula into an 
automaton, using the Spot library.

The source code is available at: \url{https://github.com/5nizza/sdf-hoa}.

\section{Experimental Results}
\label{sec:results}

We present the results of \syntcomp 2018, separated into the safety (AIGER) track and the LTL (TLSF) track. As in previous years, both tracks are separated into realizability and synthesis subtracks, and parallel and sequential execution modes.
Detailed results of the competition are also directly accessible via the web frontend of our instance of the EDACC platform at \url{http://syntcomp.cs.uni-saarland.de}.

\subsection{Safety (AIGER)-Track: Realizability}

In this track, $5$ tools competed in $12$ different configurations, $7$ in sequential execution mode and $5$ in parallel mode. We compare their performance on a selection of $234$ benchmark instances. 

We first restrict the evaluation of results to purely sequential tools, then extend it to include also the parallel versions, and finally give a brief analysis of the results.

\paragraph{Sequential Mode.}
In sequential mode, \simpleBDD competed with three configurations (basic, abs1, abs2), and the re-entered tools  each with their best-performing configuration from last year: \abssynthe (SC3 for realizability and SC2 for synthesis), \demiurge (D1real and D1synt), and \safetysynth (Basic).

The number of solved instances per configuration, as well as the number of uniquely solved instances, are given in Table~\ref{tab:results-realseq}. No tool could solve more than $165$ out of the $234$ instances, or about $70.5\%$ of the benchmark set. $24$ instances could not be solved by any tool within the timeout. 

The configurations of \lazysynt (running hors concours) solve fewer instances, but the (basic) configuration provides one unique solution, and another benchmark instance is solved by both configurations of \lazysynt, but by none of the competing tools.

\begin{table}[h]
\caption{Results: Safety (AIGER) Realizability (sequential mode only)}
\label{tab:results-realseq}
\centering
\def\arraystretch{1.3}
{\sffamily \small
\begin{tabular}{@{}llll@{}}
\toprule
Tool & (configuration)		& Solved  &  Unique \\ 
\midrule
\simpleBDD & (abs1) & 165 		& 1 \\
\simpleBDD & (basic)        & 161     & 1 \\
\safetysynth & (basic)      & 160     & 0 \\
\simpleBDD & (abs2) & 159     & 4 \\
\abssynthe & (SC3) 	 	    & 157     & 8 \\
\demiurge & (D1real)				& 125			& 18 \\
\midrule
\lazysynt & (genDel) &	98	& 0 \\
\lazysynt & (basic)	& 	94 	& 1	\\
\bottomrule
\end{tabular}
}
\end{table}

\begin{figure}
\centering
\includegraphics[width=1\linewidth]{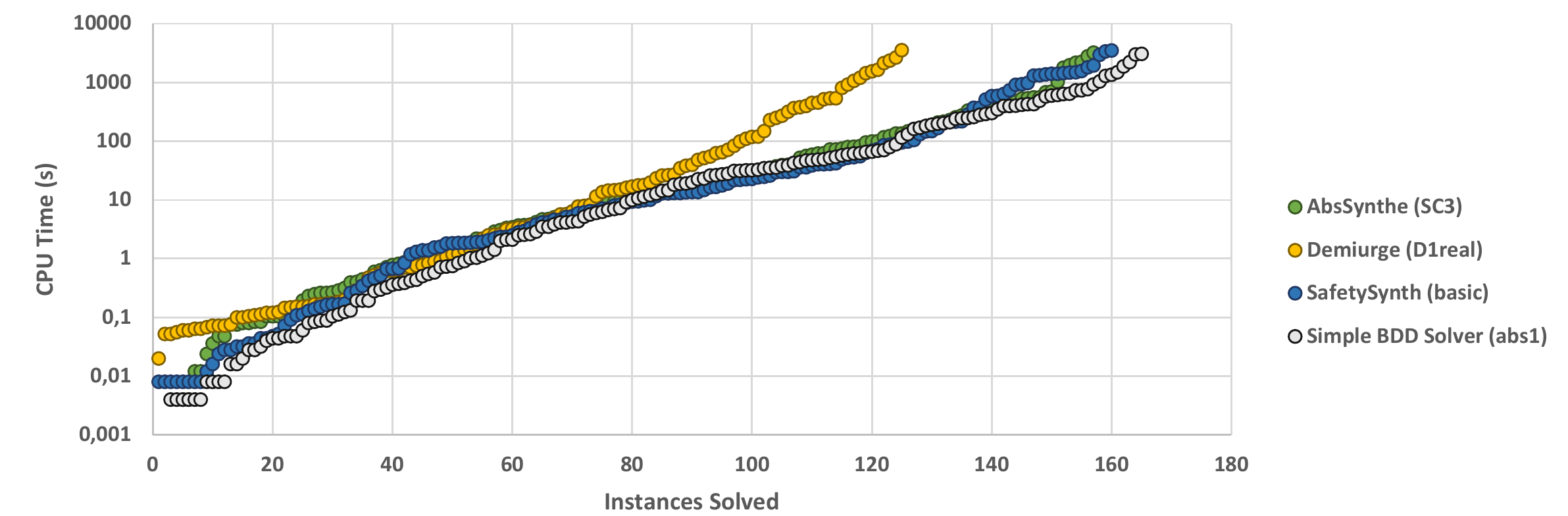}
\caption{Safety (AIGER), Runtime Cactus Plot of Best Sequential Configurations}
\label{fig:cactus-realseq}
\end{figure}

Figure~\ref{fig:cactus-realseq} gives a cactus plot for the runtimes of the best sequential configuration of each tool.
%
%The following benchmarks were solved uniquely by one tool configuration:
%\begin{itemize}
%\item \abssynthe (seq1): \texttt{moving\_obstacle\_128x128\_59glitches}
%\item \abssynthe (seq2): \texttt{mult\_bool\_matrix\_6\_6\_6, mult12}
%\item \demiurge (D1real): \texttt{beemldelec4b1\_c0to511, gb\_s2\_r2\_comp4\_REAL,\\ load\_full\_4\_comp1\_REAL, mult\_bool\_matrix\_dyn\_6\_6, very\_good\_bakery2.sym}
%\item \simpleBDD (2): \texttt{amba9match5, cycle\_sched\_12\_6\_3, cycle\_sched\_8\_7\_2,\\ driver\_a7n, driver\_b7y, driver\_c8n, factory\_assembly\_5x5\_2\_10errors,\\ factory\_assembly\_5x5\_2\_11errors, good\_bakery.false}
%\end{itemize}
%
%, including the reference implementation Aisy.

%\begin{table}
%\caption{Results: Realizability (sequential, reference solvers)}
%\label{tab:results-realseqref}
%\def\arraystretch{1.2}
%\centering
%\begin{tabular}{c|ccc}
%Tool (configuration)			& Solved \\ 
 %\hline
%%Aisy & 98
%\end{tabular}
%\end{table}

\begin{figure}
\centering
\includegraphics[width=1\linewidth]{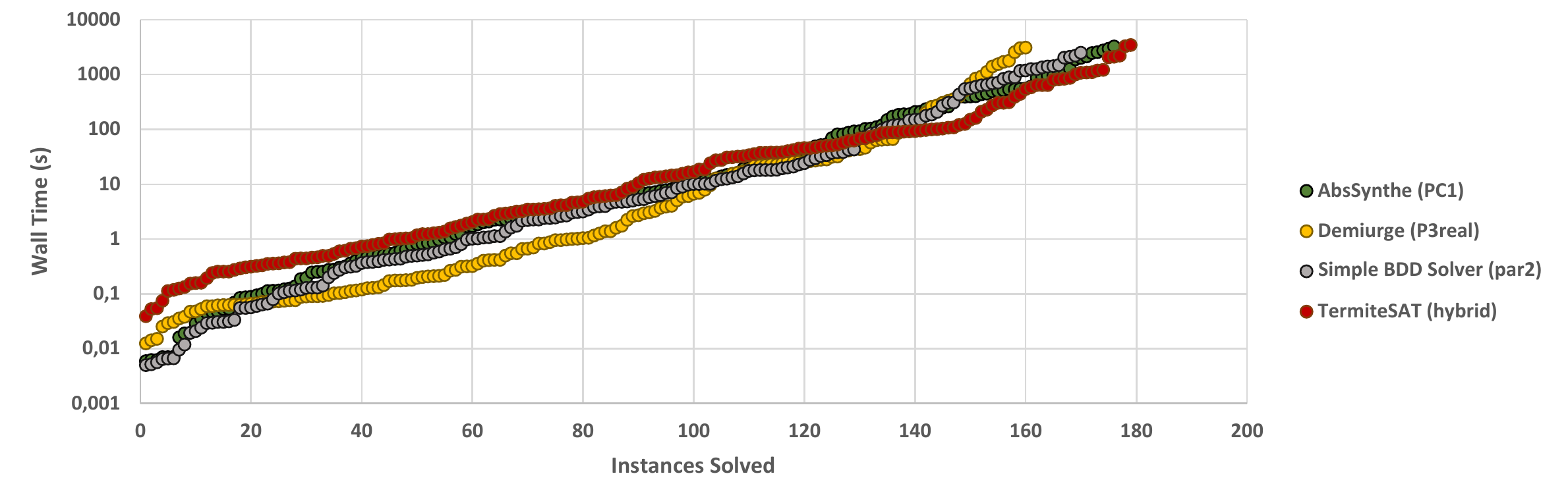}
\caption{Safety (AIGER), Runtime Cactus Plot of Best Parallel Configurations}
\label{fig:cactus-realall}
\end{figure}	

\paragraph{Parallel Mode.}
Four tools had parallel configurations for the realizability track: three portfolio configurations of \simpleBDD (par1,par2,par3), and one configuration each for the re-entered tools \abssynthe (PC1), \demiurge (P3real), and \termitesat (hybrid). They ran on the same benchmarks as in the sequential mode, but runtime of tools was measured in wall time instead of CPU time. The results are given in Table~\ref{tab:results-realpar}. Compared to sequential mode, a number of additional instances could be solved: \abssynthe, \simpleBDD and \termitesat all have one or more configurations that solve more then the best tool in sequential mode. The best result is $179$ solved instances, or about $76.5\%$ of the benchmark set. $21$ instances could not be solved by any parallel configuration, and $15$ instances could not be solved by any configuration, including the sequential ones.

\begin{table}[h]
\caption{Results: Safety (AIGER) Realizability (parallel mode only)}
\label{tab:results-realpar}
\centering
\def\arraystretch{1.3}
{\sffamily \small
\begin{tabular}{@{}llll@{}}
\toprule
Tool & (configuration)			& Solved  &  Unique \\ 
 \midrule
\termitesat & (hybrid) 		& 179 & 0 \\
\abssynthe & (PC1) 	 	 	& 176 & 1 \\
\simpleBDD & (par2)					& 170 & 0\\
\simpleBDD & (par3)					& 168 & 0\\
\simpleBDD & (par1)					& 163 & 0\\
\demiurge & (P3real) 			& 160 & 4 \\
\bottomrule
\end{tabular}
}
\end{table}

Note that in Table~\ref{tab:results-realpar} we only count a benchmark instance as uniquely solved if it is not solved by any other configuration, including the sequential configurations.

\begin{figure}[ht]
%\centering
\includegraphics[width=\linewidth]{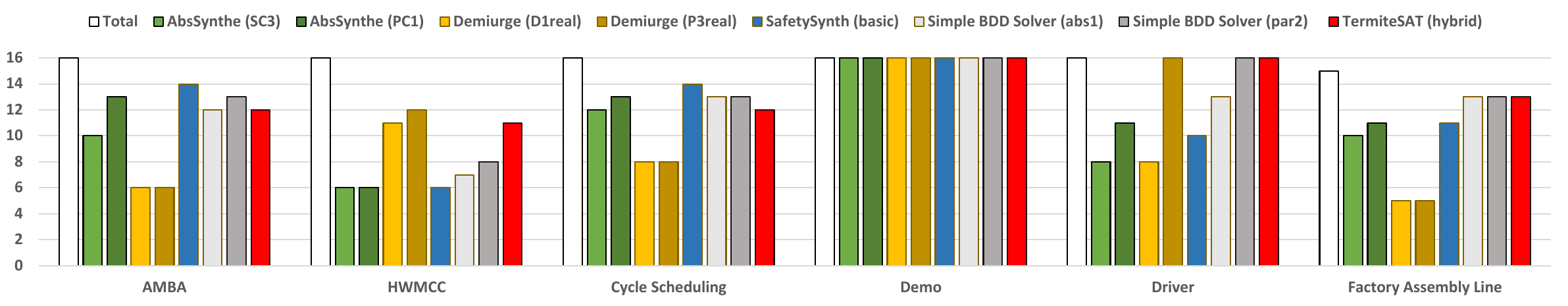}
%\caption{Realizability Track, Solved Instances by Category}
%\label{fig:bycat1}
%\end{figure}
%
%\begin{figure}[h]
%\centering
%\vspace{1em}
\includegraphics[width=\linewidth]{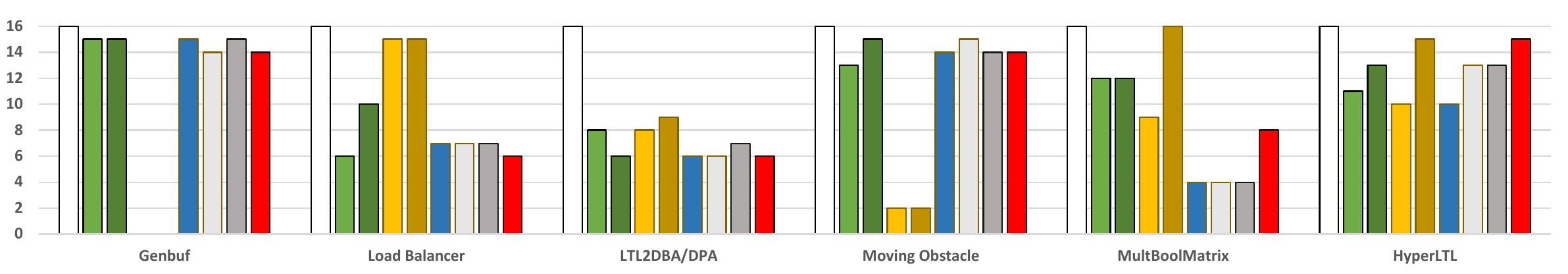}
\includegraphics[width=\linewidth]{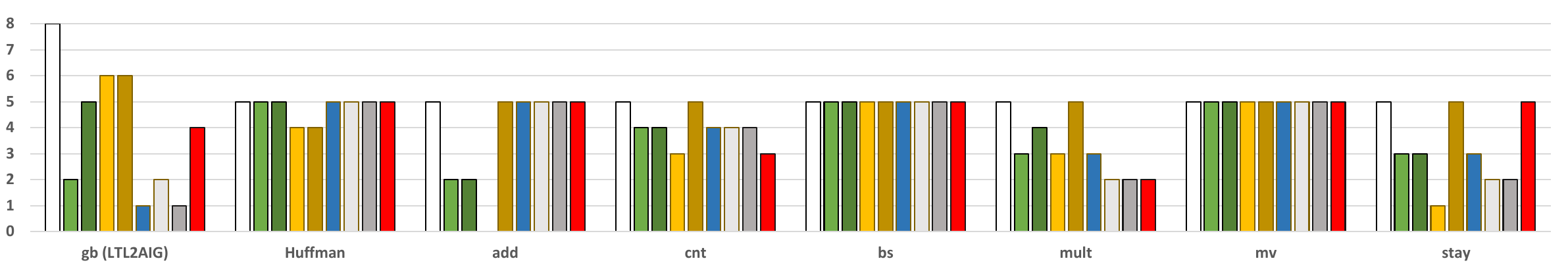}
\caption{Safety (AIGER) Realizability Track, Solved Instances by Category}
\label{fig:bycat2}
\end{figure}

\paragraph{Both modes: Solved Instances by Category.} Figure~\ref{fig:bycat2} %and \ref{fig:bycat4} 
gives an overview of the number of solved instances per configuration and category, for the best sequential and parallel configuration of each tool and different benchmark categories.

\paragraph{Analysis.}
The best sequential configuration on this year's benchmark set is again \simpleBDD (abs1), as in previous years. However, note that there are many unique solutions that are not found by this configuration, so the room for improvement should still be significant. 

In particular, \abssynthe (SC3) solves $8$ instances less than the best configuration, but also solves $8$ instances uniquely, and \demiurge (D1real) is $40$ instances behind, but can solve $18$ instances that no other sequential configuration solves. Figure~\ref{fig:bycat2} shows that \demiurge is much better than all other approaches in certain categories, like HWMCC, Load Balancer or gb (while it is far behind the other approaches in other categories, such as AMBA, Cycle Scheduling, Factory Assembly Line, and Moving Obstacle). 

Considering parallel configurations, last year's winner \termitesat (hybrid) again solves most instances, followed closely by \abssynthe (PC1), like last year. The new portfolio configurations of \simpleBDD are between $9$ and $16$ instances behind the best configuration. Finally, the parallel configuration of \demiurge again comes close to the other parallel configurations, and solves $4$ instances that no other configuration, sequential or parallel, can solve.

\subsection{Safety (AIGER)-Track: Synthesis}
In the track for synthesis from safety specifications in AIGER format, participants tried to solve all benchmarks that have been solved by at least one competing configuration in the realizability track.
Three tools entered the track: \abssynthe, \demiurge and \safetysynth, all of them in their respective best configurations for sequential and parallel mode from last year (\safetysynth does not have a separate parallel configuration).

Like last year, there are two different rankings in the synthesis track, based on the number of correct solutions and on their size compared to a reference implementation, respectively. As always, a solution for a realizable specification is only considered as correct if it can be verified (cf. Section~\ref{sec:setup}).
We present the results for the sequential configurations, followed by parallel configurations, and end with an analysis of the results.

\paragraph{Sequential Mode.}
Table~\ref{tab:results-syntseq} summarizes the experimental results, including the number of solved benchmarks, the uniquely solved instances, the number of solutions that could not be model-checked within the timeout, and the accumulated quality of solutions. No sequential configuration solved more than $154$ of the benchmarks, and $34$ instances could not be solved by any competing tool (not counting the configurations of \lazysynt that ran hors concours).

%The following benchmarks were solved uniquely by one tool configuration:
%\begin{itemize}
%\item \abssynthe (seq1):
%\item \abssynthe (seq2):
%\item \abssynthe (seq3):
%\item \demiurge (D1synt):
%\end{itemize}

\begin{table}[h]
\caption{Results: Safety (AIGER) Synthesis (sequential mode only)}
\label{tab:results-syntseq}
\centering
\def\arraystretch{1.3}
{\sffamily \small
\begin{tabular}{@{}llllll@{}}
\toprule
Tool & (configuration) & Solved & Unique & MC Timeout & Quality\\
\midrule
\safetysynth & (basic) 	& \textbf{154} & 17 & 0 & \textbf{224}\\
\abssynthe 	& (SC2) 		& 145 & 6 & 0 & 184\\
\demiurge 	& (D1synt) 	& 112 & 22 & 1 & 158\\
\midrule
\lazysynt		& (genDel)	& 94	& 0	& 4	& 137\\
\lazysynt		& (basic) 	& 91	& 2	&	4	& 125\\
%Aisy & 75 & 105 & \% & 3
\bottomrule
\end{tabular}
}
\end{table}

\paragraph{Parallel Mode.}
%In this mode, \abssynthe competed with three configurations (par1, par2, par3), and \demiurge with one configuration (P3synt). 
%In addition, we ran last year's parallel configuration of \demiurge on the new competition benchmarks.
Table~\ref{tab:results-syntpar} summarizes the experimental results, again including the number of solved benchmarks, the uniquely solved instances, the number of solutions that could not be verified within the timeout, and the accumulated quality of solutions. No tool solved more than $156$ problem instances. $24$ instances could not be solved by any parallel configuration, and $14$ instances could not be solved by any configuration, including the sequential ones.

Like in the parallel realizability track, we only consider instances as uniquely solved if they are not solved by any other configuration, including sequential ones.  Moreover, the quality ranking is computed as if the sequential configuration had also participated (which makes a difference for benchmarks that do not have a reference solution). 

%The following benchmarks were solved uniquely by one tool configuration:
%\begin{itemize}
%\item \abssynthe (par1):
%\item \demiurge (P3synt):
%\end{itemize}

\begin{table}[h]
\caption{Results: Safety (AIGER) Synthesis (parallel mode only)}
\label{tab:results-syntpar}
\centering
\def\arraystretch{1.3}
{\sffamily \small
\begin{tabular}{@{}llllll@{}}
\toprule
Tool & (configuration) & Solved & Unique & MC Timeout & Quality\\
\midrule
\abssynthe  & (PC1) 		& \textbf{156} & 0 & 0 & 204\\
\demiurge 	& (P3Synt) 	& 148 & 14 & 0 & \textbf{240}\\
\bottomrule
\end{tabular}
}
\end{table}
\paragraph{Analysis.}
As expected based on the results of previous years, \safetysynth and \abssynthe compete for the highest number of solved instances, while \demiurge again wins the quality ranking and produces a very high number of unique solutions.

\subsection{LTL (TLSF)-Track: Realizability}

In the track for realizability checking of LTL specifications in TLSF, $5$ tools competed in $7$ sequential and $3$ parallel configurations. In the following, we compare the results of these $10$ configurations on the $286$ benchmarks that were selected for \syntcomp 2018. 

Again, we first restrict our evaluation to sequential configurations, then extend it to include parallel configurations, and finally give a brief analysis.

\paragraph{Sequential Mode.}
In sequential mode, \bosy, \bowser, \party and \strix each competed with one configuration, and \ltlsynt with three configurations.

The number of solved instances per configuration, as well as the number of uniquely solved instances, are given in Table~\ref{tab:results-realseq-tlsf}. No tool could solve more than $267$ out of the $286$ instances, or about $93\%$ of the benchmark set. $14$ instances could not be solved by any of the participants within the timeout. 

\begin{table}[h]
\caption{Results: LTL Realizability (sequential mode only)}
\label{tab:results-realseq-tlsf}
\centering
\def\arraystretch{1.3}
{\sffamily \small
\begin{tabular}{@{}llll@{}}
\toprule
Tool & (configuration)		& Solved  &  Unique \\ 
\midrule
\strix					& 						& 267 & 12\\
\bosy 					& 						& 244	& 1 \\
\party 					& 			 			& 242 & 0 \\
\ltlsynt				& (ds)				& 239 & 0 \\
\ltlsynt				& (incr)			& 237 & 0 \\
\ltlsynt				& (sd)				& 233 & 0 \\
\bowser					& 						& 205	& 0 \\
\bottomrule
\end{tabular}
}
\end{table}

\begin{figure}[!hbt]
\centering
\includegraphics[width=1\linewidth]{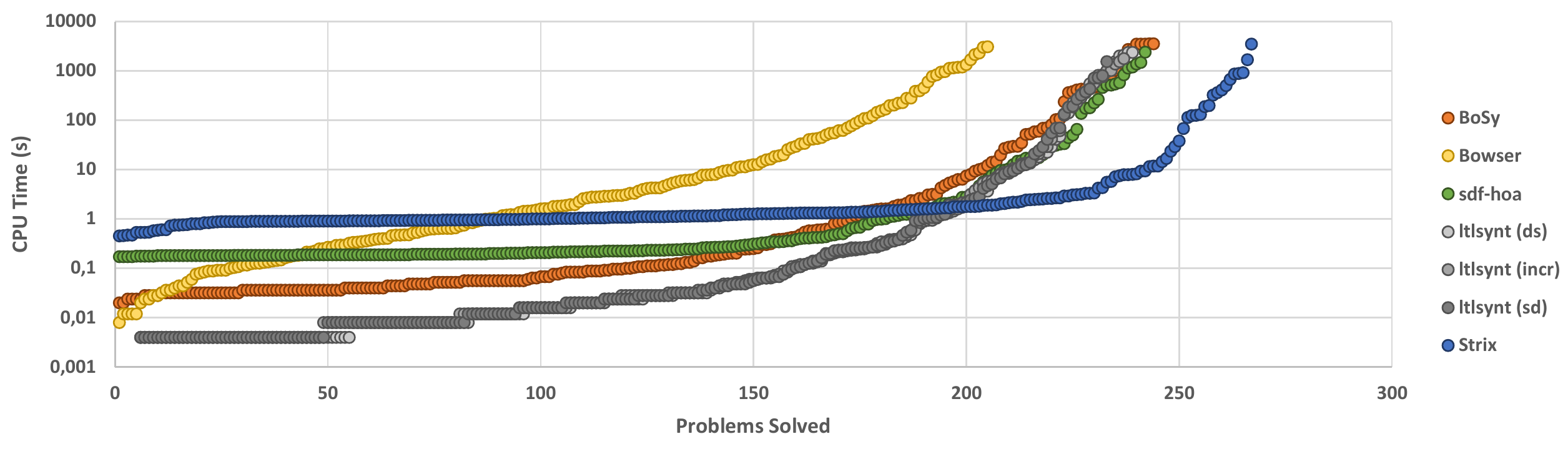}
\caption{LTL Realizability: Runtimes of Sequential Configurations}
\label{fig:cactus-realseq-tlsf}
\end{figure}

Figure~\ref{fig:cactus-realseq-tlsf} gives a cactus plot of the runtimes for all sequential algorithms in the realizability track.

%
%\begin{figure}[h]
%\centering
%\includegraphics[width=1\linewidth]{fig/TLSF-Realizability-2017-cactusall-1.pdf}
%\caption{TLSF/LTL Realizability Track: Runtimes of Parallel and Selected Sequential Configurations}
%\label{fig:cactus-realall-tlsf}
%\end{figure}	
%
%The following benchmarks were solved uniquely by one tool configuration:
%\begin{itemize}
%\item \abssynthe (seq1): \texttt{moving\_obstacle\_128x128\_59glitches}
%\item \abssynthe (seq2): \texttt{mult\_bool\_matrix\_6\_6\_6, mult12}
%\item \demiurge (D1real): \texttt{beemldelec4b1\_c0to511, gb\_s2\_r2\_comp4\_REAL,\\ load\_full\_4\_comp1\_REAL, mult\_bool\_matrix\_dyn\_6\_6, very\_good\_bakery2.sym}
%\item \simpleBDD (2): \texttt{amba9match5, cycle\_sched\_12\_6\_3, cycle\_sched\_8\_7\_2,\\ driver\_a7n, driver\_b7y, driver\_c8n, factory\_assembly\_5x5\_2\_10errors,\\ factory\_assembly\_5x5\_2\_11errors, good\_bakery.false}
%\end{itemize}
%
%, including the reference implementation Aisy.

%\begin{table}
%\caption{Results: Realizability (sequential, reference solvers)}
%\label{tab:results-realseqref}
%\def\arraystretch{1.2}
%\centering
%\begin{tabular}{c|ccc}
%Tool (configuration)			& Solved \\ 
 %\hline
%%Aisy & 98
%\end{tabular}
%\end{table}

\paragraph{Parallel Mode.}
\bosy, \bowser and \strix also entered in a parallel configuration. Again, the parallel configurations run on the same set of benchmark instances as in sequential mode, but runtime is measured in wall time instead of CPU time. The results are given in Table~\ref{tab:results-realpar-tlsf}. The best parallel configuration solves $224$ out of the $244$ instances, or about $92\%$ of the benchmark set. $10$ benchmarks have not been solved by any configuration.

\begin{table}
\caption{Results: LTL Realizability (parallel mode only)}
\label{tab:results-realpar-tlsf}
\centering
\def\arraystretch{1.3}
{\sffamily \small
\begin{tabular}{@{}llll@{}}
\toprule
Tool & (configuration)			& Solved  &  Unique \\ 
 \midrule
\strix 		& (par)						& 266			& 0\\
\bosy 		& (par)						& 242			& 0\\
\bowser		& (par)						& 212			& 0\\
\bottomrule
\end{tabular}
}
\end{table}

In Table~\ref{tab:results-realpar-tlsf}, we again only count a benchmark instance as uniquely solved if it is not solved by any other sequential or parallel configuration. Therefore, none of the parallel configurations has a uniquely solved instance. Since the parallel configurations also don't increase the number of benchmarks that can be solved, we omit a cactus plot.

\begin{figure}[!phbt]
\centering
%\vspace{3em}
\includegraphics[width=\linewidth]{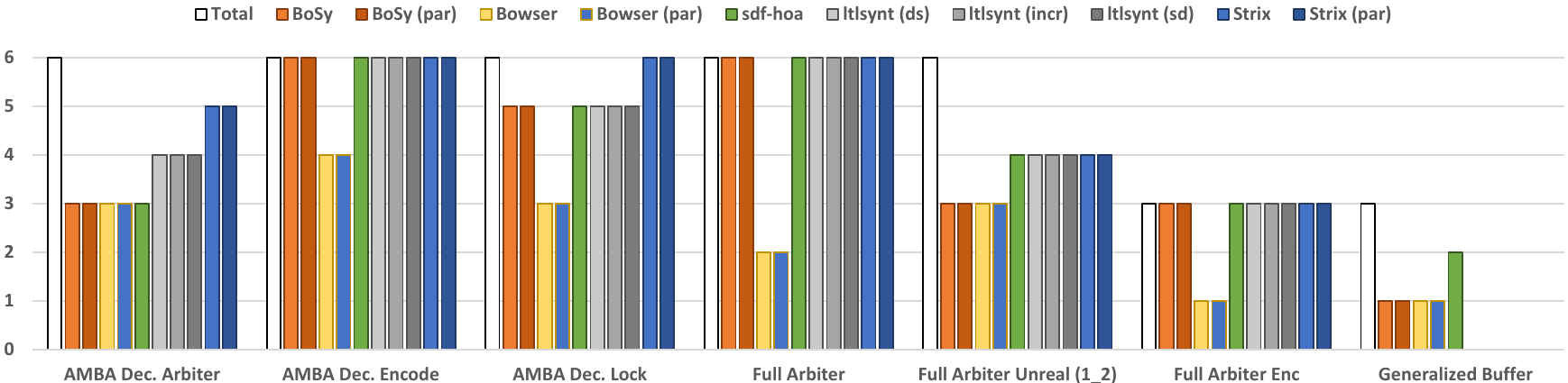}
%\caption{Realizability Track, Solved Instances by Category}
%\label{fig:bycat1}
%\end{figure}
%
%\begin{figure}[h]
%\centering
\vspace{.001em}

\includegraphics[width=\linewidth]{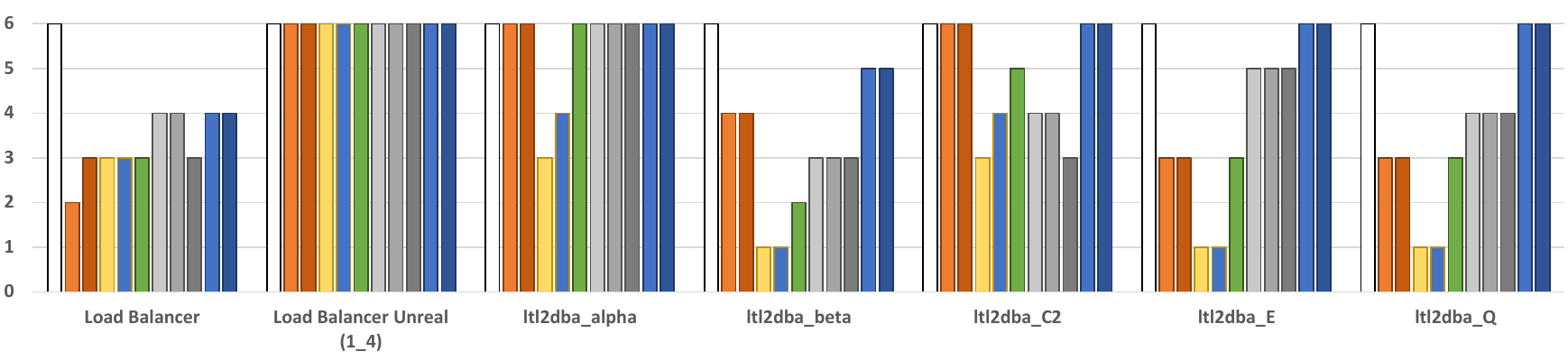}
\vspace{.001em}

\includegraphics[width=\linewidth]{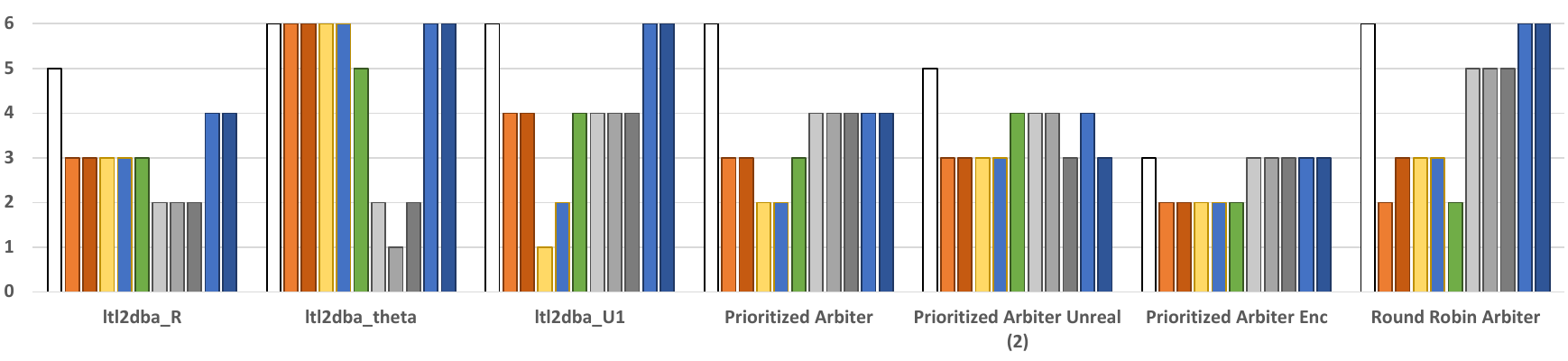}
\vspace{.001em}

\includegraphics[width=\linewidth]{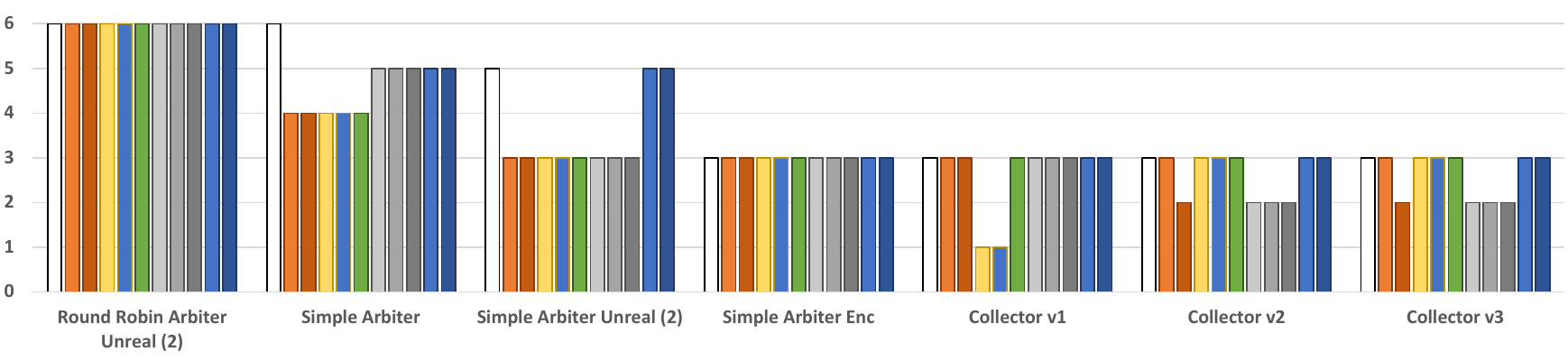}
\vspace{.001em}

\includegraphics[width=\linewidth]{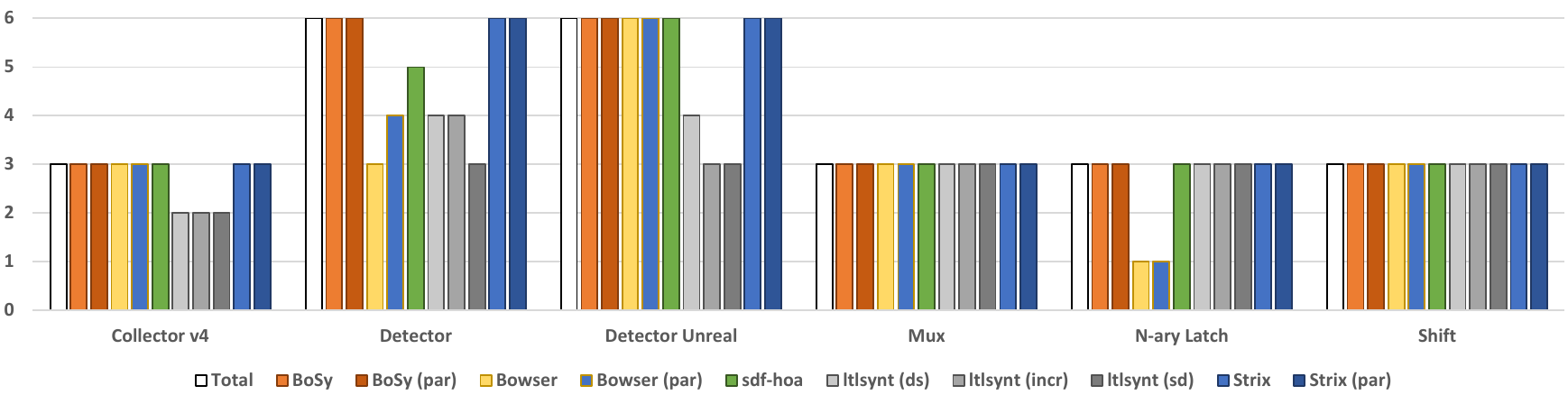}
\vspace{0.001em}
\caption{LTL Realizability: Solved Instances for Parameterized Benchmarks}
\label{fig:bycat-tlsf}
\end{figure}

\paragraph{Both modes: Solved Instances of Parameterized Benchmarks.} 
For both the sequential and the parallel configurations, Figure~\ref{fig:bycat-tlsf} gives an overview of the number of solved instances per configuration, for the $34$ parameterized benchmarks used in \syntcomp 2018.

\paragraph{Analysis.}
Like last year, the entrants of \syntcomp 2018 are very diverse: \bosy, \bowser and \party  implement different variants of bounded synthesis, and \ltlsynt and \strix use different encodings into parity games. Remarkable is the performance of the new entrant \strix, which beats by a significant margin the updated versions of the best-performing configurations from last year, \party and \ltlsynt. Moreover, \bosy has improved significantly, now solving the second highest number of instances, due to the new approach to determine realizability. 

Regarding the different configurations of \ltlsynt that are new this year, we note that the difference in the number of instances that can be solved is rather small, so the differences between these configurations do not seem to be crucial.

An analysis of the parameterized benchmarks in Figure~\ref{fig:bycat-tlsf} shows the different strengths of the approaches: \strix is almost always among the best-performing tools, with the notable exception of the \texttt{generalized\_buffer} benchmark. There, \party performs best. On a number of benchmarks, like \texttt{amba\_decomposed\_arbiter} or \texttt{ltl2dba\_E}, the bounded synthesis-based approaches are clearly outperformed by the parity game-based approaches, while on others \bosy and \party can at least beat \ltlsynt, such as for \texttt{ltl2dba\_R} or the \texttt{detector} and \texttt{detector\_unreal} benchmarks (in addition to the aforementioned \texttt{generalized\_buffer}, where they even beat \strix). 

\subsection{LTL (TLSF)-Track: Synthesis}
In the track for synthesis from LTL specifications in TLSF, participants were tested on the $272$ benchmark instances that were solved by at least one configuration in the LTL realizability track. 
Except for \party, all tools from the realizability track also competed in the synthesis track, and some of them with additional configurations: \bosy competed in two sequential and two parallel configurations, \strix in three sequential and one parallel configuration, and \bowser in three sequential and three parallel configurations. \ltlsynt competed in the same three configurations as in the realizability track. Additionally, \lazysynt participated hors concours in 

As for the safety synthesis track, there are two rankings in the LTL synthesis track, one based on the number of instances that can be solved, and the other based on the quality of solutions, measured by their size. Again, a solution for a realizable specification is only considered correct if it can be model-checked within a separate timeout of one hour (cf. Section~\ref{sec:setup}).
In the following, we first present the results for the sequential configurations, then for the parallel configurations, and finally give an analysis of the results.

\paragraph{Sequential Mode.}
Table~\ref{tab:results-syntseq-tlsf} summarizes the experimental results for the sequential configurations, including the number of solved benchmarks, the uniquely solved instances, and the number of solutions that could not be model-checked within the timeout. The last column gives the accumulated quality points over all correct solutions.

As before, the ``solved'' column gives the number of problems that have either been correctly determined unrealizable, or for which the tool has presented a solution that could be verified. Based on the $272$ instances that could be solved by at least one configuration in the realizability track, the best-performing tool solved $257$ or about $94\%$ of the benchmarks, and $11$ instances could not be solved by any tool. None of the tools provided any wrong solutions.

%The following benchmarks were solved uniquely by one tool configuration:
%\begin{itemize}
%\item \abssynthe (seq1):
%\item \abssynthe (seq2):
%\item \abssynthe (seq3):
%\item \demiurge (D1synt):
%\end{itemize}

\begin{table}[h]
\caption{Results: LTL Synthesis (sequential mode only)}
\label{tab:results-syntseq-tlsf}
\centering
\def\arraystretch{1.3}
{\sffamily \small
\begin{tabular}{@{}llllll@{}}
\toprule
Tool & (configuration) & Solved & Unique & MC Timeout & Quality\\
\midrule
\strix 					& (labels)		& \textbf{257}	& 6	& 6		& 412\\
\strix					& (min)				& 248	& 0	& 16	& \textbf{416}\\
\strix					& (basic)			& 247	& 0	& 17	& 387\\
\bosy						& (basic)			& 222	& 0 & 4		& 372\\
\ltlsynt				& (incr)			& 214	& 0	& 19	& 257\\
\ltlsynt				& (ds)				& 211	& 0	& 23	& 251\\
\ltlsynt				& (sd)				&	210	& 0	& 23	& 245\\
\bowser					& (simple)		&	206	& 0	& 0		& 308\\
\bosy						& (opt)				& 205	& 0	& 5		& 370\\
\bowser					& (synth)			& 187	& 0 & 0		& 289\\
\bowser					& (opt)				& 166	&	0 & 0		& 308\\
\bottomrule
\end{tabular}
}
\end{table}

\paragraph{Parallel Mode.}
In this mode, \bosy and \bowser competed with parallel versions of their configurations from the sequential track, and \strix competed in a portfolio approach that runs multiple configurations in parallel. 
%In addition, we ran last year's parallel configuration of \demiurge on the new competition benchmarks.
Table~\ref{tab:results-syntpar-tlsf} summarizes the experimental results, in the same format as before. No configuration solved more than $256$ problem instances, or about $94\%$ of the benchmark set. $8$ benchmarks could not be solved by any tool. None of the solutions were determined to be wrong.

As before, we only consider instances as uniquely solved if they are not solved by any other configuration, including sequential ones. Consequently, none of the solutions are unique. 
%The following benchmarks were solved uniquely by one tool configuration:
%\begin{itemize}
%\item \abssynthe (par1):
%\item \demiurge (P3synt):
%\end{itemize}

\begin{table}[h]
\caption{Results: LTL Synthesis (parallel mode only)}
\label{tab:results-syntpar-tlsf}
\centering
\def\arraystretch{1.3}
{\sffamily \small
\begin{tabular}{@{}llllll@{}}
\toprule
Tool & (configuration) & Solved & Unique & MC Timeout & Quality\\
\midrule
\strix					& (portfolio)	& \textbf{256}	&	0	&	6		& \textbf{446}\\
\bosy						& (opt,par)		& 223	& 0	&	9		&	402\\
\bosy						& (basic,par)	&	223	&	0	&	12	&	371\\
\bowser					&	(simple,par)&	212	&	0	&	0		& 315\\
\bowser					& (synth,par)	&	194	&	0	&	0		&	300\\
\bowser					& (opt,par)		& 162	&	0	&	0		&	302\\
\bottomrule
\end{tabular}
}
\end{table}

\begin{figure}[!htb]
\centering
\includegraphics[width=\linewidth]{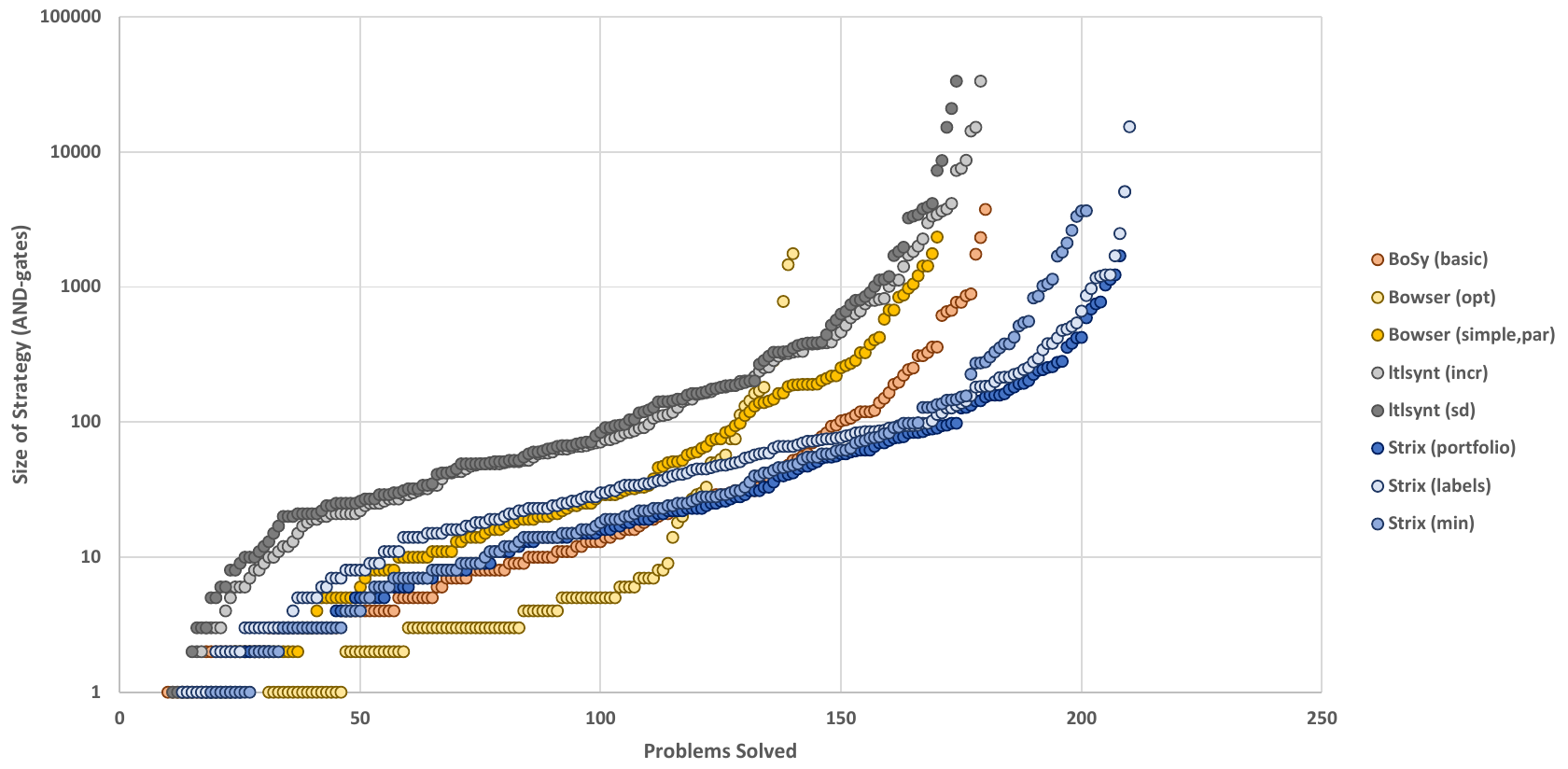}
\caption{LTL Synthesis: Solution Sizes of Selected Configurations}
\label{fig:TLSF-size1}
%\end{figure}
%
%\begin{figure}[h]
%\centering
%\includegraphics[width=\linewidth]{fig/TLSF-size2-1.pdf}
%\caption{LTL Synthesis: Solution Sizes of Parallel Configurations and \ltlsynt}
%\label{fig:TLSF-size2}
\end{figure}

\paragraph{Analysis.}

As can be expected, the number of solved instances for each tool in synthesis is closely related to the solved instances in realizability checking. The number of unique instances of \strix and \bosy decrease, in part simply due to the fact that multiple configurations of the same tool now all solve the problem that was solved uniquely before. 
Furthermore, we note that all tools except \bowser produce a significant number of solutions that could not be verified.

Considering the quality ranking, the best-performing sequential configuration is not the one that solves the highest number of instances: even though \strix (min) solves $9$ instances less, it beats \strix (labels) in the quality ranking. In parallel mode, the portfolio approach of \strix manages to combine the best of both worlds and solves just one instance less than the best sequential configuration, while significantly improving the quality score.

Figure~\ref{fig:TLSF-size1} plots the solution sizes of selected configurations. It shows, somewhat surprisingly, that many solution sizes of \strix in (min) or (portfolio) configurations are between the sizes of solutions produced by the bounded synthesis approaches implemented in \bosy and \bowser. Only configuration \bowser (opt) produces clearly smaller solutions (for roughly half of the realizable instances). In contrast, \ltlsynt in most cases produces solutions that are bigger by some margin (i.e., roughly ten times the size of the solutions of \bosy). 

A further analysis on the quality and the size of implementations shows that \bowser (opt,par) is the configuration that has the highest average quality for the problems that it does solve. Furthermore, \bosy (opt,par) and \strix (portfolio) each produce $54$ solutions that are better than the previous reference solution. The analysis for all tools is given in Table~\ref{tab:quality-synt-tlsf}.

\begin{table}[h]
\caption{LTL Synthesis: Average Quality and New Reference Solutions}
\label{tab:quality-synt-tlsf}
\centering
\def\arraystretch{1.3}
{\sffamily \small
\begin{tabular}{@{}llll@{}}
\toprule
Tool & (configuration) & Avg. Quality & New Ref. Solutions \\
\midrule
\bowser					& (opt,par)		& \textbf{1.862} & 34\\
\bowser					& (opt)				& 1.848 & 33\\
\bosy						& (opt,par)		& 1.802 & \textbf{54}\\
\bosy						& (opt)				&	1.794 & 48\\
\strix					& (portfolio)	& 1.741	& \textbf{54}\\
\strix					& (min)				& 1.666 & 49\\
\bosy						& (basic) \& (basic,par) & 1.661	& 39\\
\strix					& (labels)		& 1.595	& 45\\
\strix					& (basic)			& 1.555	& 46\\
\bowser					& (synth,par)	& 1.544	& 3\\
\bowser					& (synth)			& 1.534	& 2\\
\bowser					& (simple) \& (simple,par) & 1.484 & 4\\
\ltlsynt				& (incr)			& 1.206	& 8\\
\ltlsynt				& (ds)				& 1.173	& 7\\
\ltlsynt				& (sd)				& 1.171	& 8\\
\bottomrule
\end{tabular}
}
\end{table}

\section{Conclusions}
\label{sec:conclusions}

SYNTCOMP 2018 was the fifth iteration of the reactive synthesis competition, and showed that there is still substantial progress from year to year. While the safety track saw only minor updates and one new tool that was entered hors concours, the LTL track again saw major changes, including the new tool \strix that won all official categories of the track.

In addition, also the benchmark set of the LTL track has grown significantly, and this year for the first time we did not release all benchmarks beforehand.

{ \small
\myparagraph{Acknowledgments}
The organization of \syntcomp 2018 was supported by the Austrian Science Fund
(FWF) through project RiSE (S11406-N23) and by the German
Research Foundation (DFG) through project ``Automatic Synthesis of 
Distributed and
Parameterized Systems'' (JA 2357/2-1), and its setup and execution by the 
European Research Council (ERC) Grant OSARES (No.~683300). 

%The development of \abssynthe was supported by an F.R.S.-FNRS and FWA fellowships, and
%the ERC inVEST (279499) project.

%The development of \demiurge was supported by the FWF through project RiSE
%(S11406-N23, S11408-N23).

The development of \safetysynth and \bosy was supported by the 
ERC Grant OSARES (No.~683300). 

The development of \strix was supported by the German Research Foundation (DFG) projects ``Game-based Synthesis for Industrial Automation'' (253384115) and ``Verified Model Checkers'' (317422601) and by the ERC Advanced Grant PaVeS (No.~787367).

%The development of \simpleBDD was supported by a gift from the Intel
%Corporation.
%
%NICTA is funded by the Australian Government through the
%Department of Communications and the Australian Research Council through the ICT
%Centre of Excellence Program.
}

%\nocite{*}
\bibliographystyle{eptcs}
\bibliography{synthesis}%,references}
\end{document}